\documentclass[12pt]{iopart}
\usepackage[utf8]{inputenc}
\usepackage[dvipdf,pdftex]{graphicx}

\begin{document}


\title[Frequency Stabilized DBR Diode Laser for Dimensional Metrology]
{Laser Source for Dimensional Metrology: Investigation of an Iodine Stabilized System Based on Narrow Linewidth 633 nm DBR Diode}

\author{
	Simon Rerucha$^1$, 
	Andrew Yacoot$^2$,
	Tuan M. Pham$^1$, 
	Martin Cizek$^1$, 
	Vaclav Hucl$^1$, 
	Josef Lazar$^1$ 
	and Ondrej Cip$^1$}

\address{$^1$ Institute of Scientific Instruments of the Czech Academy of Sciences (ISI), Kralovopolska 147, 612 64 Brno, CZ}
\address{$^2$ National Physical Laboratory (NPL), Hampton Road, Teddington, Middlesex, TW11 0LW, UK}

\ead{res@isibrno.cz}

\begin{abstract}
	
We demonstrated that an iodine stabilized Distributed Bragg Reflector (DBR) diode based laser system lasing at a wavelength in close proximity to $\lambda = 633\,$nm could be used as an alternative laser source to the He-Ne lasers in both scientific and industrial metrology.
This yields additional advantages besides the optical frequency stability and coherence: inherent traceability, wider optical frequency tuning range, higher output power and high frequency modulation capability. 
We experimentally investigated the characteristics of the laser source in two major steps: 
first using a wavelength meter referenced to a frequency comb controlled with a hydrogen maser 
and then on a interferometric optical bench testbed where we compared the performance of the laser system with that of a traditional frequency stabilized He-Ne laser.
The results indicate that DBR diode laser system provides a good laser source for applications in dimensional (nano)metrology, especially in conjunction with novel interferometric detection methods exploiting high frequency modulation or multiaxis measurement systems.
\end{abstract}



\submitto{\MST}
\maketitle



\section{Introduction}

Helium-Neon lasers (He-Ne) working at $632.99\,$nm vacuum wavelength are popular sources of laser radiation for length metrology. 
They are used as a light source for interferometers for precision measurement of distances in electron-beam writers, atomic force microscopes or in dimensional measuring systems in an industrial environment, to name but a few applications  \cite{dai2006accurate,lazar2009local}. 
These gas lasers gained their popularity due to narrow spectral emission line (down to several kHz), superb Gaussian beam profile, a stable operation of the output optical power, a turn-key operation and ease of use thanks to a visible wavelength that make them ideal for precise adjustment of interferometric setups. 
The technology of gas tube production is well developed thanks to more than $50$-years history so that these lasers offer long service life together with high operational reliability.

On the other hand, the typical commercially available frequency stabilized single mode He-Ne lasers have also several disadvantages. 
The low output power of a few hundreds of microwatts prevents splitting the laser beam to provide a light source for several interferometers simultaneously.
The mode-hop free tuning range of their optical frequency is limited to only a few hundreds of MHz. 
Since the tuning employs the control of the laser tube temperature (and e.g. the piezo-driven models are not widely available), the bandwidth response of the tuning is also low. 

The emergence of laser diodes has led to compact lasers with large tuning range and high power output.  
Effort has been directed towards replacing conventional gas laser with diode lasers, but the problem was that, generally, the linewidth was much wider and narrowing it posed a technological challenge (in terms of designing and manufacturing the active layer and the waveguide). 
The most promising alternative has been the Extended Cavity Laser setups (ECL) that have been used to build frequency standards \cite{simonsen1997iodine, edwards1999iodine}.
 
This arrangement enabled narrowing the spectral width of the emission line down to tens of kHz while preserving the large tuning range. For example, an iodine stabilized ECL laser at $632.9\,$nm (vacuum wavelength) with approximately ten times better relative stability than iodine stabilized He-Ne laser has been reported \cite{zarka2000international}. 
Unfortunately, the mechanical construction of the external resonator and a short lifetime of the laser diode with an anti-reflective coating on the output facet render the ECL systems difficult to operate reliably on a long term scale.

Recent progress in laser diode manufacturing enabled the production of Distributed Bragg Reflector (DBR) laser diodes lasing at $633\,$nm, i.e. at a very close proximity to the traditional wavelength band which is the most widely used wavelength for metrology. This opens the possibility for alternative laser systems for metrology that feature a comparable performance in terms of the spectral linewidth, the frequency stability and noise figures and brings the additional benefits such as the inherent traceability, wide-range mode-hop free optical frequency tunability, high-bandwidth tunability, higher output power and operational reliability. 
The tuning options could be exploited for the novel detection techniques in interferometry that requires a source with frequency modulated beam\cite{Sensors12b} and enables the simplified optical arrangement in the measurement systems.
Additionally, since many off the shelf optical components are optimised for $633\,$nm, there is a wide variety of those components readily available.

In this paper we present a concept and an experimental investigation of the performance of a laser system based on a DBR laser diode. 
The diode, lasing at $633\,$nm, is frequency stabilized onto a molecular iodine I$_2$ transition using linear spectroscopy. 
We have assembled the setup which comprises the laser, beam shaping and collimation and fibre-coupling optics, an iodine cell and related electronics. 
The frequency properties of the laser system were investigated using a dedicated wavelength meter  
while auxiliary measurement was being carried out using a polarimeter.
We have also assembled the experimental optical bench test-bed interferometric setup that was used to 
characterize the properties of the laser system in a typical interferometric application: stability and noise of the measured displacement. 

These properties were experimentally compared to those of a traditional frequency stabilized He-Ne laser in order to assess the usability of the DBR diode laser in typical applications in the field of length metrology.  
It is important to point out that the system is intended as a reliably operable laser system for both the scientific and industrial metrology rather than a fundamental laboratory frequency standard.

The rest of this paper is organized as follows: 
The Methodology section presents the overview of the laser system setup, the stabilization procedure and the interferometric optical bench test-bed.
The Experimentation section describes the design of the experiments and presents the experimental results of the laser source's frequency and polarization stability and the noise properties investigated by the means of experimental interferometric measurement.
The Discussion section presents the interpretation of the results and discusses the necessary context. 
Finally the section Conclusion concludes the paper.

\section{Methodology}

Stabilization of the optical frequency of a laser diode is a principal prerequisite for its use as a light source for an optical interferometer. The optical frequency of the laser diode, which typically has a wide tuning range, is essentially determined by two parameters: the injection current and the operating temperature of the semiconductor chip. Changes of these parameters significantly influence the refractive index of the semiconductor active layer and the optical resonator length itself. Therefore an accurate control of  the laser diode optical frequency requires the use of an extremely low noise source for the injection current control as well as a fast and sensitive temperature controller. Both of these controllers are required to have a high dynamic range in order to maintain the optical frequency of the laser diode output accurately enough against the reference absorption medium (the molecular iodine in a glass cell in our case). Another important task is the management and conditioning of the laser beam since the beam delivery from the laser head to the interferometer is frequently via fibre feed.

\subsection{DBR diode laser setup}

\begin{figure}[htbp]
	\centering
	\includegraphics[width=.9\textwidth]{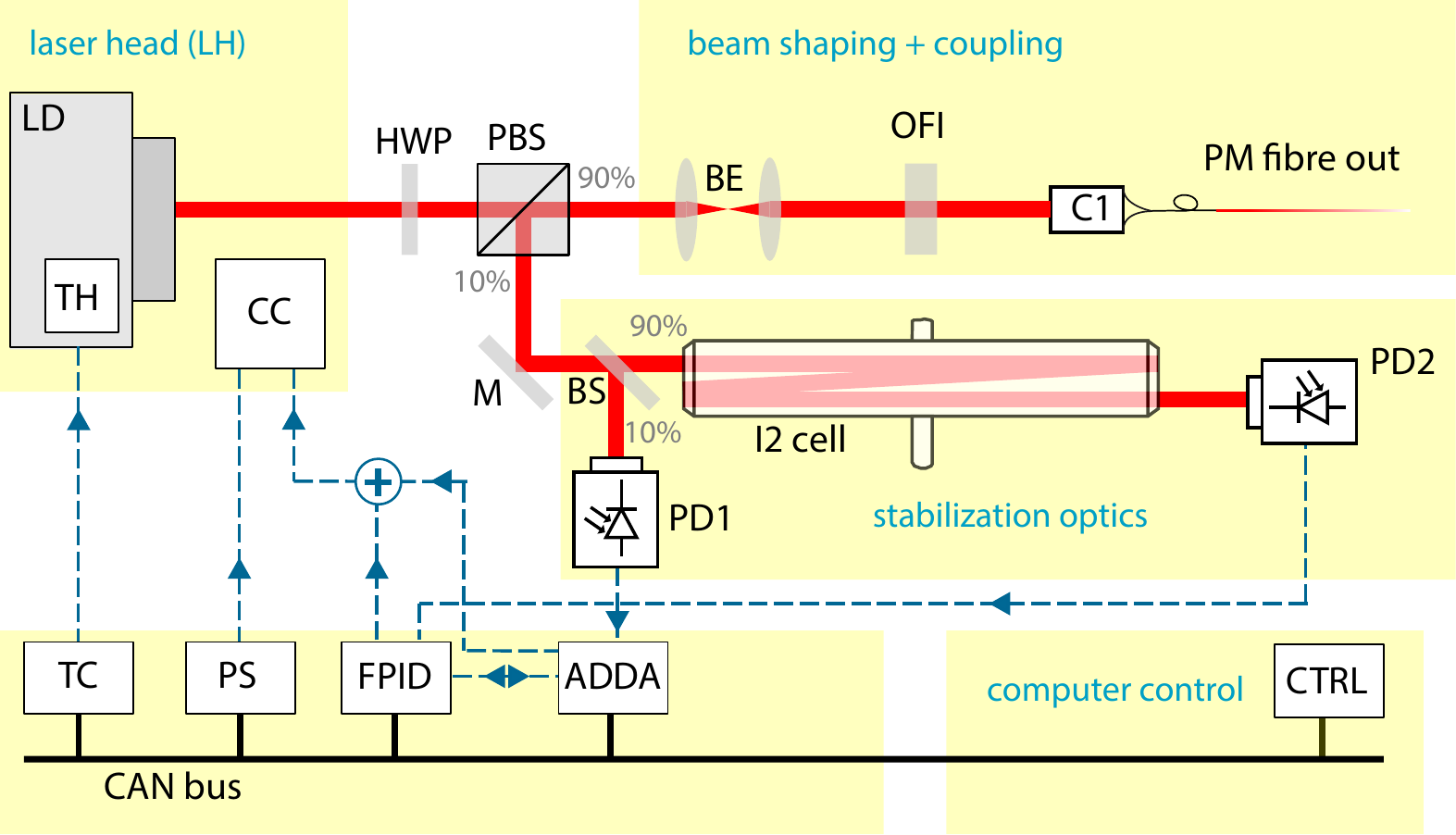}
	\caption{Overview of the laser system setup: the laser head (LH) contains the laser diode (LD) with integrated  thermal control elements (TH) and current controller (CC);  the control electronics comprising temperature controller (TC), low-noise power supply (PS), fast analogue PID controler (FPID) and analog-to-digital and digital-to-analog control module (ADDA); all of those controlled from computer software (CTRL); optical system incorporates beam expander (BE), optical Faraday isolator (OFI), fibre coupling collimator (C1), half-wave plate (HWP), polarizing beamsplitter (PBS), mirror(M), beam sampler (BS), photodetectors (PD1, PD2) and iodine cell (I2 CELL).}
	\label{f_exp}
\end{figure}

A schematic diagram of the set up is shown in Figure \ref{f_exp}. The heart of the laser system is the Distributed Bragg Reflector (DBR) free-space laser diode (LD; EYP-DBR-0633-00010-2000-TOC03-0000, by EagleYard \cite{eye2013}) which is manufactured in a standard TO-3 package. 
The output beam, propagating through a small window on the top of the laser diode package, is focused by an aspheric lens into a collimated beam with rectangular profile ($2\,$mm $\times\ 6\,$mm). 
It passes through a zero-order half-wave plate (HWP) that, in conjunction with a polarizing beamsplitter (PBS), splits the beam between the fibre coupling branch ($\approx 90$~\% of intensity) and the optical frequency stabilization branch (remaining $10$~\%).

In the coupling branch the beam is collimated by a pair of lenses (BE) that effectively form a telescope. The optical Faraday isolator (OFI) is used to prevent unwanted reflection from further optics falling back on the laser diode chip.  After the OFI the beam is fibre-coupled using aspheric collimator into a polarization maintaining (PM) fibre. 

In the stabilization branch the beam is further split by a $10:90$ beam sampler (BS) The output with $90\,$\% part passes through the iodine gas cell (I2 CELL) and is sensed by a photodetector (PD1) while the remaining $10\,$\% part is directly monitored by another photodetector (PD2).

The temperature controller (TC) in conjunction with the thermal control elements (TH; comprising a  thermistor and a Peltier cell) built into the laser diode package are used to control the diode chip temperature. 
The injection current is driven by a current controller (CC; built-in to the LH) that allows for externally driven fast-bandwidth current changes, i.e. for modulation and stabilization. 
The CC relies on an ultra low-noise power supply (PS). Both the TC and CC controllers (\cite{lazar1997electronics, mikel2002, lazar2003cc}) are microprocessor-based PID regulators and custom-built at ISI. 
The TC provides a resolution of $0.2\,$mK corresponding to an optical frequency tuning resolution of $6.75\,$MHz (corresponding to $0.009\,$pm in the wavelength) and a theoretical dynamic range of $104\,$dB. 
The current controller achieves a resolution of $53\,$nA, i.e. $40\,$kHz of optical frequency of the laser diode ($0.053\,$fm in the wavelength) and a theoretical dynamic range of $98\,$dB. 

The signal from PD2 serves as input for the analog PID controller (FPID) that drives the diode injection current via the CC. 
Note that the analog controller has bandwidth up to $3\,$MHz that matches the frequency response of the diode current modulation circuitry. 
The signals from PD1 together with the monitoring outputs from FPID are sampled by analog to digital converters of the digital signal controller based analog control module (ADDA). 
The digital to analog converters on the same module drive the operational parameters for FPID and CC, represented by analog voltages.

The useful optical power at the fibre output of the experimental system  is $2.5\,$mW compared to nominal $10\,$mW diode output. This rather low value is due to beam astigmatism and slightly inhomogeneous beam output from the diode that prevented an efficient coupling ratio. The particular insertion losses were as follows: $0.5\,$dB the optical Faraday isolator, $0.5\,$dB the collimation optics (telescope lenses), $4.5\,$dB at the fibre coupling ($\approx 35\,$\% efficiency), $0.5\,$dB for stabilization branch of the laser system. The total insertion losses for the useful output of the laser diode are $6\,$dB.

%
%

\subsection{Laser diode optical frequency stabilization}

As mentioned earlier, the laser system utilizes the absorption lines of molecular iodine as an absolute frequency reference \cite{wallard1972} for optical frequency control of the laser diode. 
A modification of the linear spectroscopy technique \cite{petru1993design} is used, i.e. a portion of optical intensity is passed through the glass cell filled with molecular iodine vapour. 
When the wavelength of the passing beam coincides with that of a particular atomic transition, absorption of the beam energy occurs and the photodiode behind the cell observes the decrease of incident light intensity. 
The iodine cell used (I2 CELL), also custom made at ISI, is $30\,$cm long and filled to $14\,^{\circ}$C saturation temperature \cite{lazar2009,hrabina2014}. The wedged ($0.5^{\circ}$) windows with anti-reflective coating are partly covered with a dielectric reflective coating so that multiple passes are possible without external mirrors.

Traditionally, the first harmonic detection technique \cite{wallard1972} is used to find the bottom of a selected absorption line. 
The laser wavelength is slightly frequency modulated (by injection current in our case) such that the rate of absorption is sensed by the photodiode within the small wavelength range as given by the modulation depth. 
The first derivative of the photodiode signal PD2, retrieved by the synchronous demodulation, is then proportional to the error signal for the frequency regulator. 
One of the principal drawbacks is that the optical frequency of the laser has to be modulated. 
This requires either additional modulation element (e.g. AOM or EOM) or the modulation of either the diode chip temperature or the injection current. 
The latter inevitably makes the useful output also frequency modulated. 
Within the course of development, we have implemented and tested this well-proven technique, but for the purposes of the experimentation presented in this paper we have developed and used a slightly modified approach:

To mitigate these effects, we have employed a stabilization technique that locks the optical frequency not onto the center of the Doppler-broadened transition line, but onto a particular point (set-point) on its side. This approach uses the direct intensity signal after the cell as the regulation input for the lock-in control loop, so that no modulation of the optical frequency is necessary. The set-point used for locking is determined as the point on the transition line side with minimal (or maximal) first derivative, i.e. with the steepest slope and thus the most distinctive frequency discrimination (in the terms of the frequency to intensity response). 

Naturally, the stabilization onto the absolute value of intensity signal makes the technique susceptible to amplitude fluctuations. 
These are compensated with a second order regulation control loop. 
This loop compensates the intensity changes induced by the frequency control: 
it tunes the diode temperature so that the injection current is kept in a narrow range -- this approach exploits the fact that the temperature changes cause a smaller intensity change than the injection current change. 

As noted before, the stabilization is carried out by the pair of standalone control modules: ADDA and FPID. 
These modules implement the first-order regulation (analogue) and provide the interface for the second order regulation (digital), respectively. 
All the control parameters of the laser diode such as the injection current and operating temperature set-points together with the regulation parameters are controlled from a dedicated LabView application through the CAN communication bus that interconnects all electronic modules and the computer.

Unless stated otherwise, the laser has been stabilized onto an iodine transition at $\nu = 473.138\,$THz ($\lambda = 633.625\,$nm in terms of vacuum wavelength) \cite{salami2005}. 
The fundamental uncertainty associated with the stabilization of the optical frequency is derived from the quantization step of temperature control ($6.5\,$MHz) and the injection current control ($40\,$ kHz).

%
%

\subsection{Interferometric optical bench testbed}

\begin{figure}[htbp]
	\centering
	\includegraphics[width=.8\textwidth]{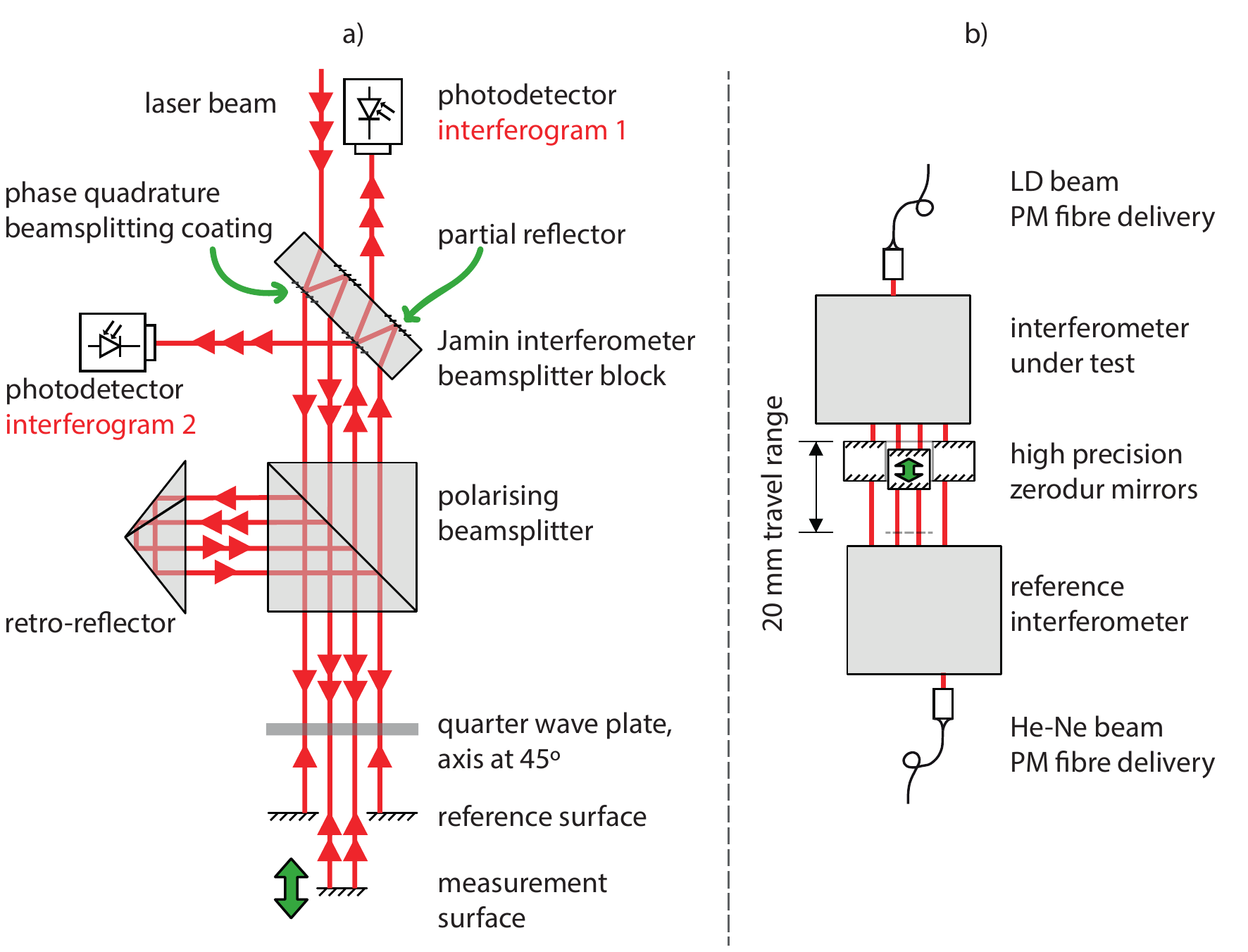}
	\caption{a) NPL Plane Mirror Differential Optical Interferometer schematic \cite{yacoot2000}; b)~experimental setup arrangement with two interferometers}
	\label{f_jamin}
\end{figure}

The interferometric testbed setup is a vital part of instrumentation that enabled a comparison of the newly developed stabilized DBR diode based laser and a traditional He-Ne laser (dual polarization mode thermally stabilized) working at $632.8\,$nm wavelength for use with dimensional metrology.
The interferometric measurements have been performed using the NPL Plane Mirror Differential Optical Interferometer (see Figure \ref{f_jamin} for schematics \cite{yacoot2000,pisani2012comparison}) and electronics developed at NPL. 
The optical arrangement is differential, based on Jamin beamsplitter with dielectric coatings to achieve the phase shift between the two beams. The homodyne detection of the interference phase uses two photodetectors for quadrature signal output (Sine and Cosine) and an additional reference photodetector that samples overall intensity of the incident beam for the further compensation of the intensity and polarization fluctuations.  The interferometer has a balanced length of both arms so that the wavelength stability of the input laser does not significantly increase the total noise properties of the interferometer and the interferometer is less sensitive to local variations in refractive index.

The signals from photodetectors are processed by dedicated analog electronics comprising two modules: a preamplifier unit and a conditioning unit. The first unit converts the photodiode signals to voltages while the latter normalizes the amplitudes relatively to the intensity reference and scales them so as to produce two signals in approximate quadrature \cite{birch1990}.
The signals, at this point representing the Sine and Cosine with an amplitude of $\pm9\,$V, are then low pass filtered (a second order passive RC filter, $f_c = 15\,$kHz) and sampled by a FPGA-fitted data acquisition card (DAQ, model PCI 7833R by National Instruments) with analog to digital converters (ADC) that have a 16-bit resolution over the nominal $\pm10\,$V range (practically the range is slightly extended). 
Besides sampling, the FPGA performs an on-line Heydemann non-linearity correction \cite{heydemann1981} and phase unwrapping. Dedicated control/acquisition software is used for the data logging and experimentation control.

According to error propagation laws, the quantization step of the phase detection $\delta\Phi$ is given as 
\begin{eqnarray}
\delta\Phi &=& \frac{1}{x^2+y^2}\sqrt{y^2\delta x^2 + x^2\delta y^2},
\end{eqnarray}
where $x,y$ are Sine/Cosine immediate signal magnitudes and $\delta x, \delta y$ is the quantization step of the ADC.
Considering that $\delta x = \delta y$, we can simplify to
\begin{eqnarray}
\delta\Phi &=& \frac{1}{x^2+y^2}\sqrt{y^2\delta x^2 + x^2\delta x^2}\nonumber\\
		  &=& \frac{1}{x^2+y^2}\delta x (\sqrt{y^2 + x^2})	 \nonumber\\
		  &=& \frac{\delta x}{\sqrt{x^2+y^2}}
\end{eqnarray}
 Note that $\sqrt{x^2+y^2}$ is effectively equal to the radius of the Lissajou figure.  
The resulting $\delta\Phi$ is expressed in radians. Knowing that $\delta x = \delta y = 0.35 \times 10^{-3}\,$V, given by the output code bit width (16~bits) to input voltage range ratio ($-11.4\,$V to $11.6\,$V), then $\delta\Phi = 3.51 \times 10^{-4} / 9  = 3.90 \times 10^{-5}\,$ radians. 
With the double-pass design for $633\,$nm wavelength, the quantization resolution is equal to $0.98\,$pm. 

Considering manufacturer-declared relative accuracy of the DAQ to be $2.17\,$mV for a single point measurement and $0.22\,$mV when averaged with $n = 100$, the uncertainty associated with the resolution and properties of the analog to digital conversion is $6.8\,$pm or $0.7\,$pm, respectively. 
The manufacturer-declared absolute accuracy is to be $7.78\,$mV, which would suggest a contribution to the uncertainty of $24.5\,$pm. However, we can assume that the systematic error could be compensated by the Heydemann correction implemented within the detection chain.

%
%

\section{Experimentation}

We have carried out several sets of measurements in order to verify the fundamental parameters of the laser system as well as its feasibility for use in dimensional nanometrology. 
At the beginning we presented the observed polarization and intensity beam properties as well as the tunability range of the LD.  
Then we have conducted experiments using a typical interferometric setup for displacement measurement that gave us an overview of the achievable displacement resolution and showed the real impact of frequency (in)stability and noise. 

In several experiments we have employed two lasers for a comparison. The “device under test” laser was the newly developed DBR diode laser setup (referred to as ‘LD’) and for the reference we have used a frequency stabilized He-Ne laser  (by Research Electro-Optics, further referred to as the  ‘He-Ne’).

%
%

\subsection{Intensity dependence and polarizarion stability}

Although the interferometer’s working principle is based on the use of dielectric coatings to achieve phase quadrature rather than polarization optics, a polarizing beamsplitter and quarter wave plate together with a  corner cube are used to laterally shift the beams after first the reflections from reference- and measurement mirrors. 
If the polarization of the input laser beam changes, the resulting magnitude of interference fringes can vary resulting in scale non-linearity of the interferometer, that could be nonetheless compensated. 
Therefore we have inspected the state of polarization (SOP) in the terms of polarization rotation, degree of polarization (DOP) and the intensity of the laser beam output from the free-running diode (i.e. without fibre coupling and frequency stabilization). The data were measured by a polarimeter (PAX5710 by Thorlabs), that had accuracy as follows: SOP angle $\pm0.25\,$\%,  DOP $\pm0.5\,$\% and relative intensity change  $\pm0.3\,$dB . 

\begin{figure}[htbp]
	\centering
	\includegraphics[width=\textwidth]{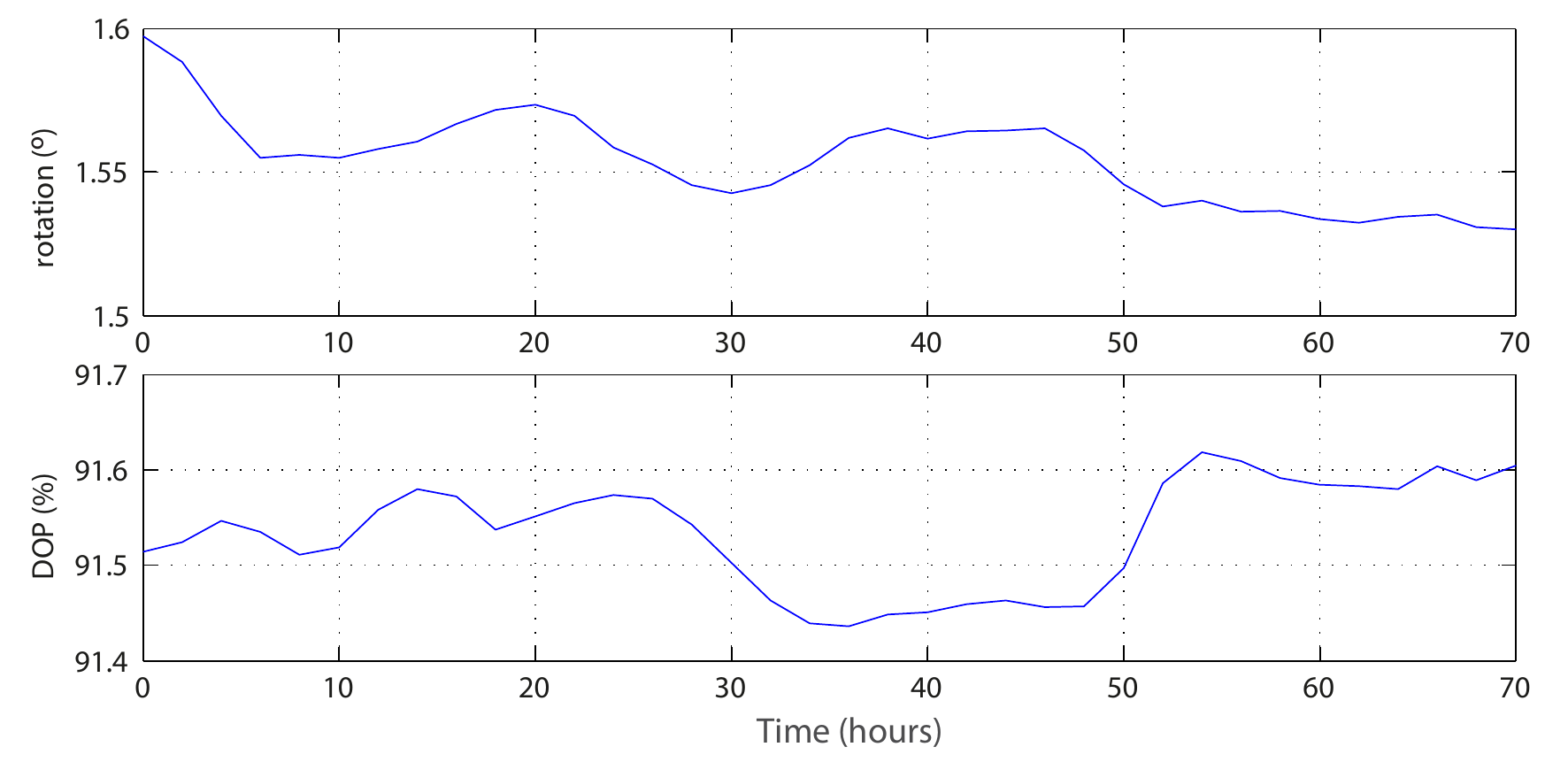}
	\caption{The beam parameters stability in time: state of polarization (top) and degree of polarization (bottom).}
	\label{f_polTime}
\end{figure}

Figure \ref{f_polTime} shows the SOP fluctuating within $\pm0.02^{\circ}$ and DOP drifting within~$\pm0.08$\,\% over the period of 70 hours. The intensity (not displayed) fluctuated within $0.01$ \%.  

\begin{figure}[htbp]
	\centering
	\includegraphics[width=\textwidth]{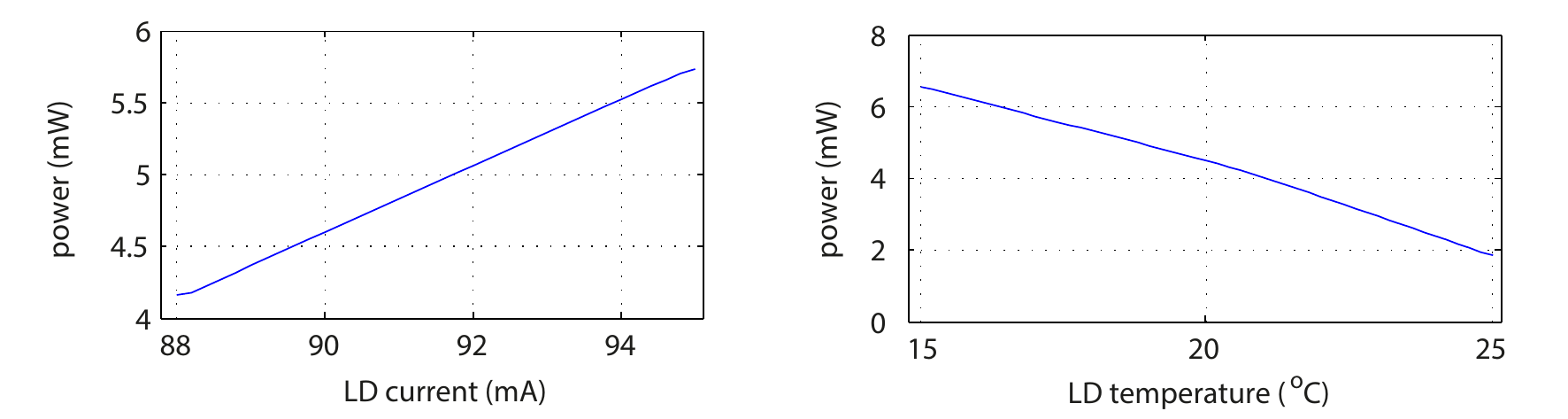}
	\caption{The output power of the DBR diode laser depending on LD temperature (right) and injection current (left).}
	\label{f_polTuning}
\end{figure}

Figure \ref{f_polTuning} shows the output power depending on the diode chip temperature and injection current. The temperature tuning revealed that the slope of intensity growth was $-0.47\,$mW$\,\cdot\,$K$^{-1}$. The injection current modulation displayed an output power rise by $0.24\,$mW$\,\cdot\,$mA$^{-1}$.

The important outcome from these measurement was the information that both the state and degree of polarization fluctuate only to a limited extent and there are no sudden changes in long-term operation. Similarly, the output power has a smooth dependence on the tuning parameters without any hops or distortions.

Note that the influence of the polarization and intensity fluctuation on practical measurements is strongly dependent on the particular application. 
All the characteristics mentioned here are subject to a significant influence once the laser beam is fibre-coupled or any polarization optics, such as a wave plate or polarization discriminator, is inserted. Nonetheless, these fluctuations would contribute to amplitude fluctuations and noise.

%
%

\subsection{Frequency drifts and tunability}

\begin{figure}[htbp]
	\centering
	\includegraphics[width=\textwidth]{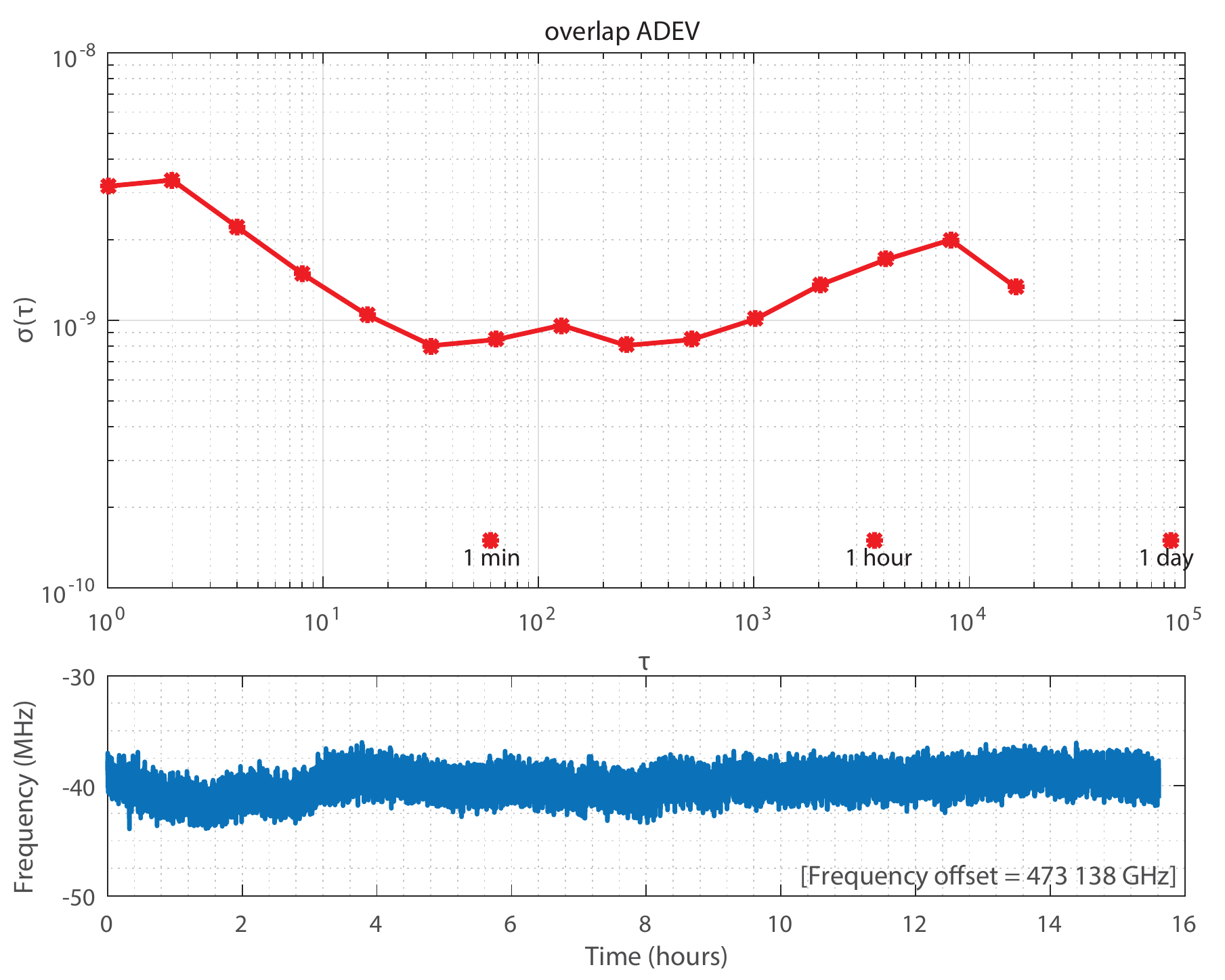}
	\caption{The Allan deviations (top) and corresponding recording of frequency drifts transition, as measured with the wavelength meter (bottom)}
	\label{f_drifts}
\end{figure}

We have measured the frequency stability and tunability using a wavelength meter (WSU10-IR by HiFinesse, absolute accuracy $100\,$MHz).  To secure the traceability to a fundamental standard, the wavelength meter used an auxiliary reference represented by a $729\,$nm laser that was referenced via phase lock to a selected component of a frequency stabilized optical frequency comb \cite{cizek2014twostage} locked to hydrogen maser.
Figure \ref{f_drifts} shows the recording of wavelength drift relative to the $633.625\,$nm I$_2$ absorption transition and the corresponding Allan's deviations. 
The observed relative frequency stability  was $0.85 \times 10^{-9}$ over $1$ minute and $1.65 \times 10^{-9}$ over $1$ hour. To assess the repeatability of the frequency locking, we have repeatedly tuned the laser optical frequency away by $>10\,$GHz, switched it off and on and stabilized the laser again.
The observed reproducibility was $\sigma = 7.1\,$MHz for $n = 7$.

\begin{figure}[htbp]
	\centering
	\includegraphics[width=\textwidth]{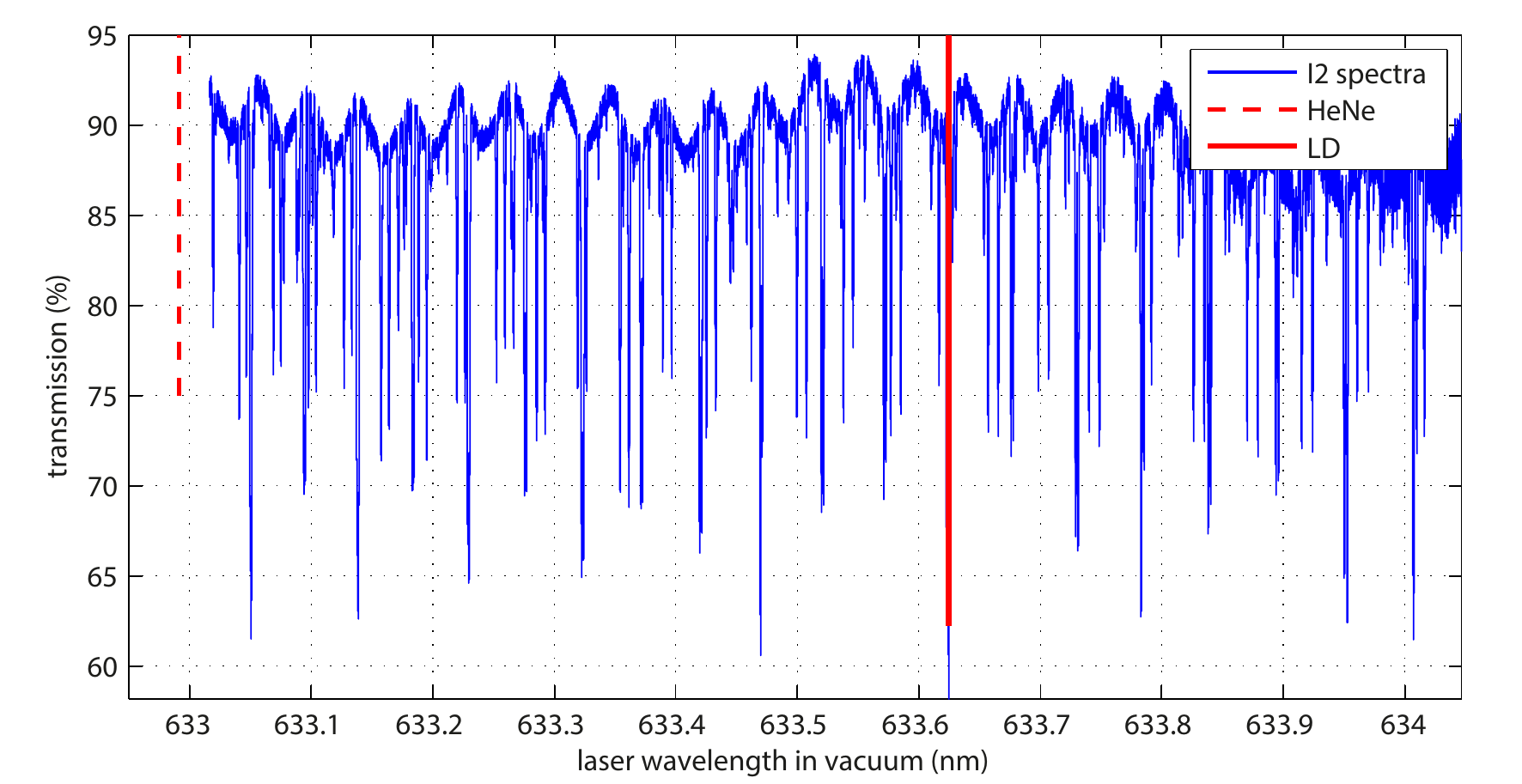}
	\caption{Wide-band modulation demonstrated on the record of large scan of an iodine absorption spectra as measured by the newly designed laser system. The vertical dashed line indicated the position of He-Ne laser’s wavelength ($632.991\,$nm vacuum wavelength) respective to the tuning wavelength of the DBR diode laser. The vertical solid line indicates the transition ($633.625\,$nm vacuum wavelength) used for LD locking}
	\label{f_iodine}
\end{figure}

The declared frequency tuning range of the laser diode is $470\,$pm (corresponding to $344\,$GHz) within the recommended operational range of the temperature tuning. However, by exceeding the range recommended by manufacturer (while still keeping within the absolute maximum ratings), we were able to achieve the $1\,$nm mode-hop free tuning range. Figure \ref{f_iodine} shows the observed iodine absorption spectra in the range from $633.017\,$nm to $634.045\,$nm (from $472.823\,$THz to $473.592\,$ THz). We have tested the tunability range of the DBR diode laser system at $I_{\mathrm{set}} = 110\,$mA and in the temperature range from $5\,^{\circ}$C to $29.5\,^{\circ}$C. 
 
%
%
\subsection{Displacement stability and noise investigation}

The experiments were carried out with the two interferometers counter-aligned opposite to each other (as indicated in Figure \ref{f_jamin}b). 
This arrangement allowed us to compensate for the influence of the stage’s mechanical instabilities.

We measured the displacement of the measuring mirror as it was moved through the measurement range $D$ from $0\,$mm to $20\,$mm.
The translation was provided by a slip-stick translation stage (SmarAct SLC-1730). 
At the start of the travel range (referred to as $D = 0\,$mm) the optical path difference was close to a balanced state where the optical path difference was minimal for both the interferometers. 
The measurement of stability was made in discrete steps, where the stage was turned off to allow it to stabilize in order to minimise mechanical drifts (these could not be totally eliminated).  
We have also measured the temperature, relative humidity and atmospheric pressure with a custom unit\cite{hucl2013automatic} for the compensation of fluctuations in the the refractive index of air (RIA).

First we converted the fringe measurements from both interferometers into nanometres using the known centre wavelength of both lasers and a factor for the cosine alignment error. %
Then we subtracted the scales to get their difference that basically express the coincidence of the displacement measurement between the two interferometers.

\begin{figure}[htbp]
	\centering
	\includegraphics[width=\textwidth]{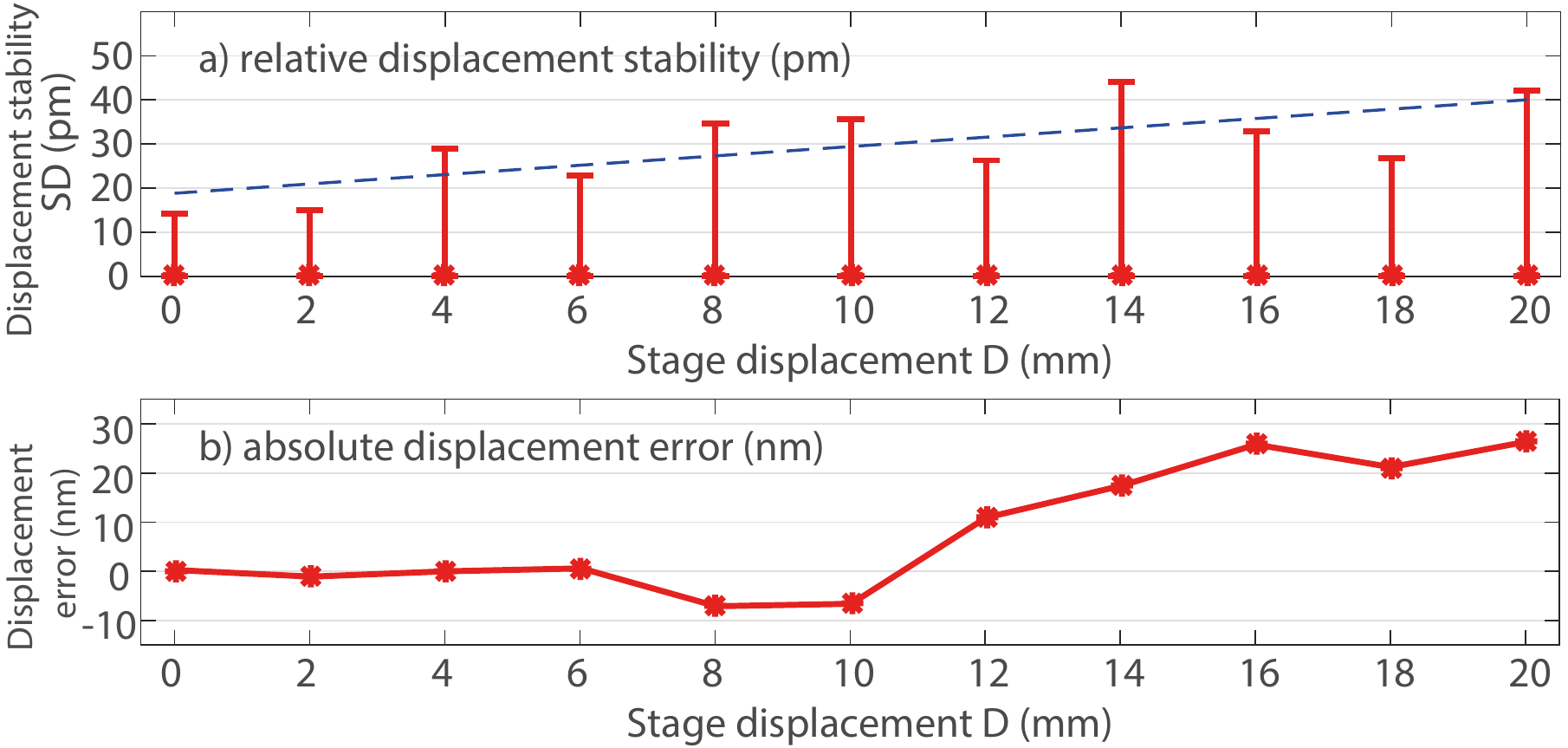}
	\caption{a) Displacement measurement stability $\sigma$ over $60$ seconds with the bandwidth from $0\,$kHz to $20\,$kHz. Dashed line indicates the contribution of the laser induced noise to displacement, which grows with increased optical path difference; b)~Mean displacement coincidence at individual stage positions}
	\label{f_stab}
\end{figure}

Figure \ref{f_stab}a displays the observed one-sigma relative displacement stability, evaluated at each measurement point through the travel range with the bandwidth of $20\,$kHz. 
We see that the coincidence is between $15\,$pm and $50\,$pm. 
The trend, shown by the dashed line, indicates the instability increase of $1.06\,$pm$\,\cdot\,$mm$^{-1}$ and therefore an approximate uncertainty contribution in order of $2.5 \times 10^{-9}$. 

Figure \ref{f_stab}b shows the average absolute displacement error at individual measurement points with mean error of $7,74\,$nm and the standard deviation of $12,77\,$nm.  
It is apparent that the magnitude of this displacement error ($1.46\,$nm$\,\cdot\,$mm$^{-1}$, indicated by the dashed lines) is significantly higher than the stability observed at the individual points. 
This discrepancy suggests that the displacement errors are caused by other factors than the frequency stability of the lasers and that the combined uncertainty of $1.46 \times 10^{-6}$ is significantly contributed to by the mechanical factors such as translation stage guidance errors.  

\begin{figure}[htbp]
	\centering
	\includegraphics[width=\textwidth]{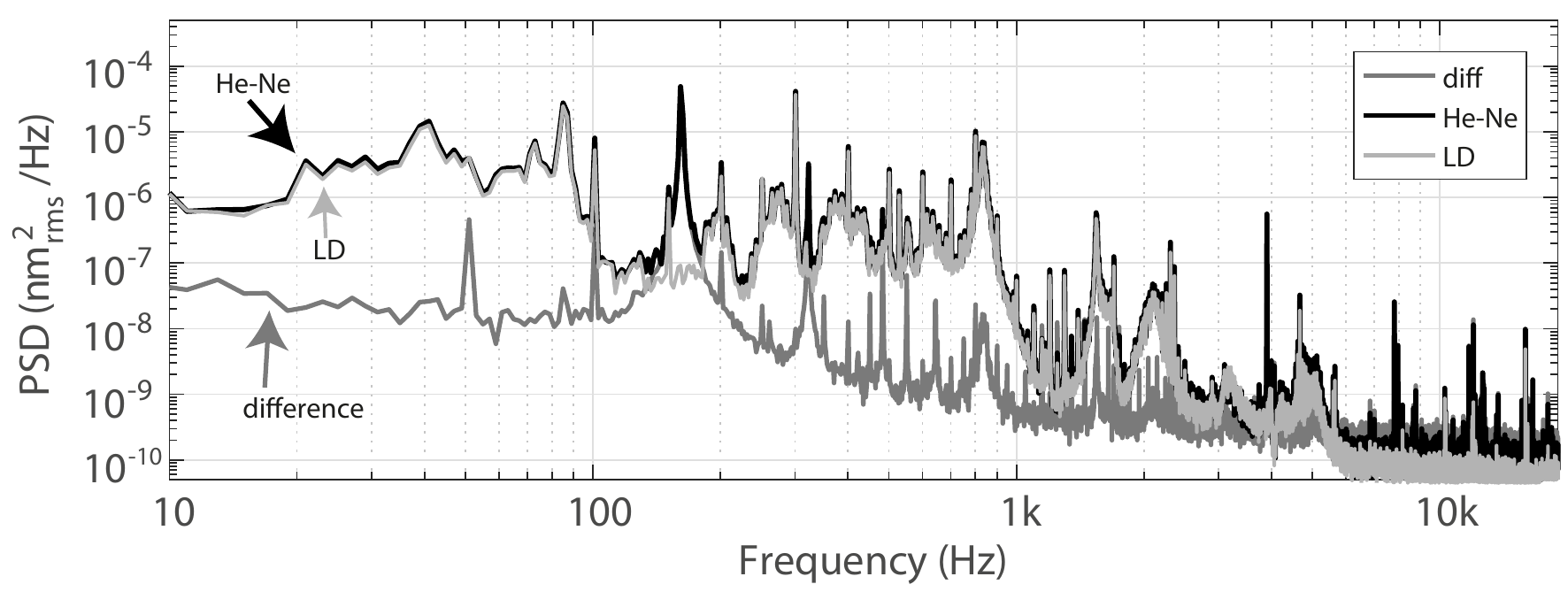}
	\caption{Baseline noise spectra at the nearly balanced state, averaged ($n = 20$), showing the power spectral density (PSD) of the noise in both lasers and the corresponding spectra obtained from the difference between the two interferometers’ scales}
	\label{f_floor}
\end{figure}

A more detailed insight of what are the contributions to the displacement noise is provided by an analysis in a frequency domain. Figure \ref{f_floor} shows the power spectral density (PSD) of displacement measurement in the balanced position. 
There is a very tight coincidence between the two lasers while the noise amplitudes of their difference are significantly lower. 
The noise in the range up to $1\,$kHz can be attributed to the mechanical and environmental factors (i.e. drifting of the translation stage and RIA fluctuations) that are mutually compensated -- the significant exceptions are the $150\,$Hz in the He-Ne laser and the $50\,$Hz and $100\,$Hz multiples. 

\begin{figure}[htbp]
	\centering
	\includegraphics[width=\textwidth]{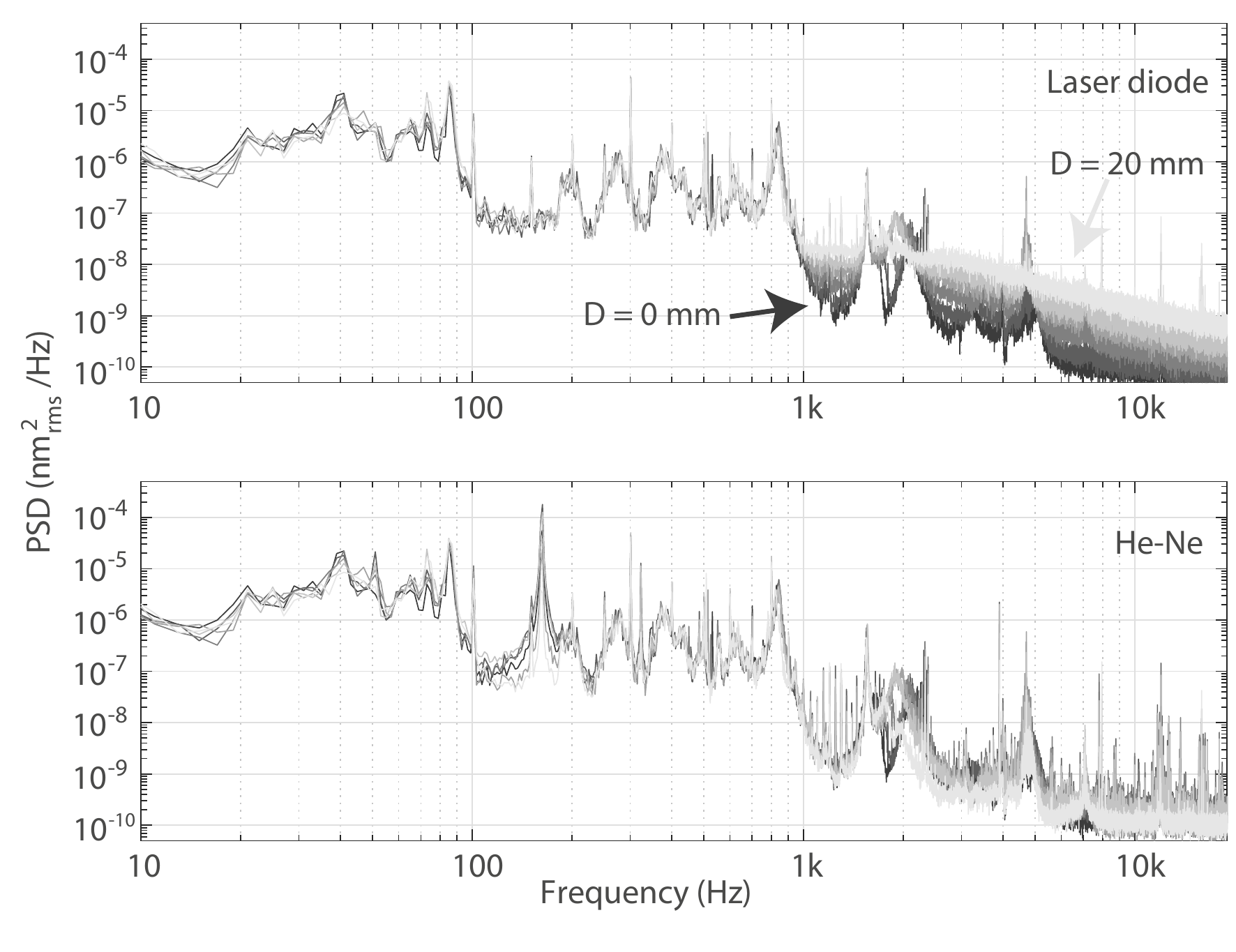}
	\caption{Noise spectra depending on the travel range $D$: the laser induced noise remains constant with the He-Ne while slightly growing in the frequency range above  $1\,$kHz with the diode laser}
	\label{f_laserNoise}
\end{figure}

The noise above $1\,$kHz can be attributed to the lasers' frequency noise. 
Figure \ref{f_laserNoise} shows the displacement noise measured with both lasers: the LD laser (Figure \ref{f_laserNoise}a) and the reference He-Ne (Figure \ref{f_laserNoise}b). 
The darkest shade shows the noise at $D=0\,$mm and the lightest at $D=20\,$mm.
The increase in the LD laser noise with the displacement in the range above $1\,$kHz can be probably attributed to the its wider linewidth. 
In contrast, with the He-Ne the noise remains constant in the entire frequency range. 
There is a small decrease above $1\,$kHz which is probably related to the fact that one of the interferometer’s dead paths is actually getting shorter with the increasing stage displacement.

\begin{figure}[htbp]
	\centering
	\includegraphics[width=\textwidth]{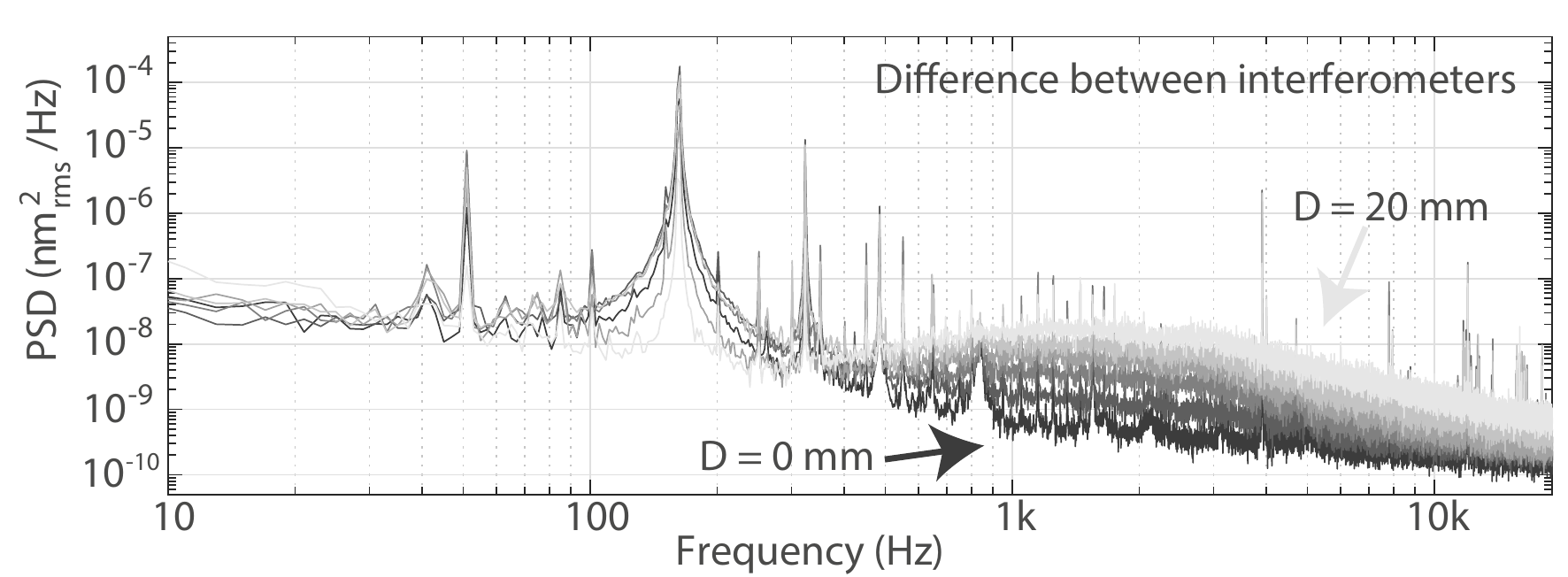}
	\caption{Noise spectra calculated from the displacement deviation between the two interferometers: there is an apparent increase between $0.4\,$kHz and $10\,$kHz}
	\label{f_diffNoise}
\end{figure}

\begin{figure}[htbp]
	\centering
	\includegraphics[width=.9\textwidth]{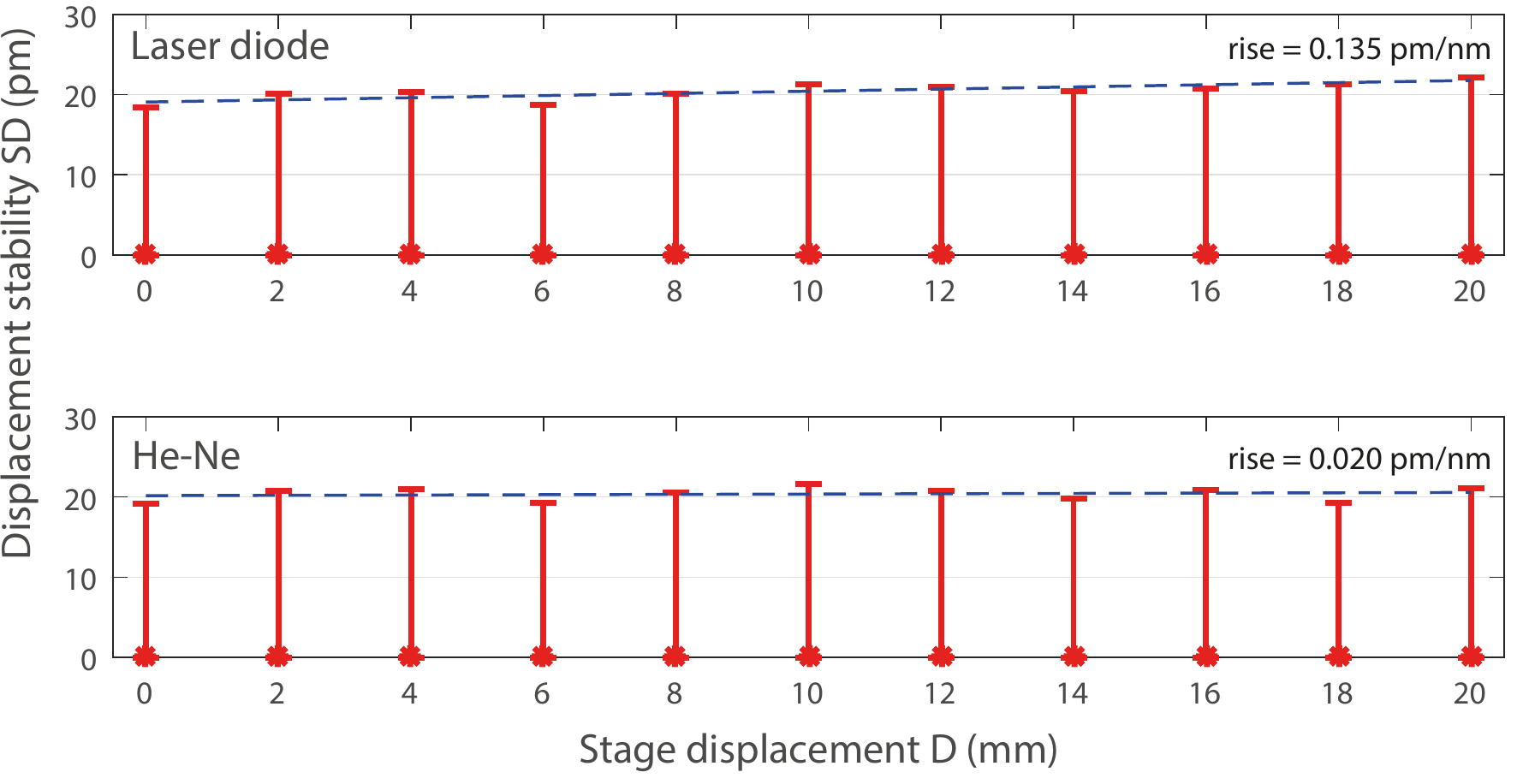}
	\caption{Contribution of stability $\sigma$  with He-Ne (top) and LD (bottom) over $60$ seconds with the bandwidth from $400\,$Hz to $20\,$kHz. Dashed line indicates the contribution of the laser induced noise to displacement, which grows with increased optical path difference}
	\label{f_stabXY}
\end{figure}

The influence of the increasing noise in the LD laser is also visible in the spectra for the displacement difference, shown in Figure \ref{f_diffNoise}, in the range above $0.4\,$kHz. 
To inspect the total contribution of the noise from this range, we applied a high-pass filter (f$_c = 0.4\,$kHz) onto the displacement signals from the two interferometers and compared the contribution total noise amplitudes with respect to the stage translation (see Figure \ref{f_stabXY}). The contribution is slightly higher with the LD  ($0.14\,$pm$\,\cdot\,$mm$^{-1}$) than with the He-Ne ($0.02\,$pm$\,\cdot\,$mm$^{-1}$) and could be probably attributed to frequency noise of the laser.
Nonetheless the contribution is generally negligible compared to the total noise, especially below $0.5\,$kHz which can be attributed to the mechanical vibrations and instability of the optical testbench and the refractive index of air fluctuations.

A theoretical drift of $1\,$MHz in the optical frequency of the laser source would cause a wavelength shift of $2.1\,$pm per $1\,$mm of the optical path difference in the interferometer, so with four-pass interferometric system such a drift will introduce $8.4\,$pm difference in observed phase per $1\,$mm of mirror displacement. Consequently we can assume that the contribution of the laser frequency noise is not a significant contribution to the uncertainty of the interferometric measurement system.

%
%

\section{Discussion}

The results indicate that the DBR diode laser based system (LD) is a good alternative to the stabilized He-Ne laser as a light source for interferometers in dimensional metrology. Nonetheless there is a variety of aspects that need to be considered. 

The spectral linewidth of the laser diode is declared by the manufacturer to be less than $1\,$MHz \cite{eye2013}, implying the coherence length of approximately $92\,$m. 
Experimental observation indicates that the actual linewidth is approximately $1.8\,$MHz in the free running mode and $1.2\,$MHz ($80\,$m coherence length) when locked onto the line side. The spatiotemporal coherence parameters of the DBR diode laser system are worse when compared to those of frequency stabilized He-Ne lasers, but they do not represent a significant contribution to uncertainty.  

The degree of polarization (DOP) is significantly worse but most of the interferometric setups incorporate some type of polarizer with high polarization extinction ratio such as a Glan-Thompson or Glan-Taylor prism that improves the DOP at the expense of small additional insertion loss. The residual amplitude noise of the laser beam passed through the crystal can be compensated with appropriate circuitry in the quadrature signal detection unit. In this case the DOP does not significantly contribute to the total displacement noise.

Some benefits of the diode laser system we present are the mode-hop free frequency tuning options. The temperature tuning enables wideband tuning over almost $0.5\,$nm (in the recommended range of operational parameters). 
On the contrary the modulation of the injection current allows frequency modulation in smaller range but with significantly higher bandwidth (up to MHz rates).
Both tuning options nonetheless cause a residual amplitude modulation. 
The modulation options can be useful for traditional applications (e.g. for frequency modulation spectroscopy) but they also open the way for novel detection methods, based on optical frequency modulation \cite{Sensors12b}, that require significantly less complex optical arrangement in comparison with the homodyne systems and just a single frequency laser in comparison with heterodyne systems.

The stabilization of the DBR diode laser optical frequency using the linear spectroscopy of iodine vapour with first-harmonic locking or the side line locking is not the state of the art technique in terms of the achievable frequency stability that would be demanded e.g. in the field of fundamental optical frequency standards. It nonetheless represents a robust, stable and repeatable mechanism that ensures the defined traceability to fundamental properties. 
In conjunction with the wavelength tunability options the approach provides a reference frame for locking at arbitrary absolute frequencies along the tuning range which can be obviously an interesting framework e.g. for wavelength sweeping absolute displacement interferometry \cite{mikel2005absolute} or the aforementioned detection methods.
It is important to note, that the side-line locking scheme has been used as a proof-of-principle due to the intended investigation based on the interferometric setup with homodyne detection. For the intended use with the frequency modulation based detection techniques, the frequency modulation needed for the first harmonic detection could be simultaneously used for the fringe counting technique.

Another significant advantage of the DBR diode laser is the nominal output power, that could be approximately seven-times higher than in a single longitunidal mode He-Ne laser. We failed to achieve this high intensity due to poor coupling ratio between some of the optics and the strong sensitivity of the diode to backreflections. There is also a fibre coupled version of the DBR laser diode from the same manufacturer (EYP-DBR-0633-00005-2000-BFY02-0000 by EagleYard), but the preliminary tests showed that the available samples exhibited even stronger sensitivity to backreflections and didn’t offer such an extended wavelength tuning range without mode hops \cite{tuan2015}. 
Currently, the manufacturer provides a new version of the narrow linewidth DBR diode lasing at $633\,$nm that has an integrated free-space beam shaping/collimation optics. Since they specify up to $80\,$\% coupling efficiency during laboratory tests, this option will be considered for future development of DBR diode based laser systems. 

The DBR diode laser possibly represents an interesting alternative to He-Ne’s in terms of costs. The most cost-critical parts are the laser diode itself and the iodine cell. While at the time of writing the cost of the DBR diode is approximately comparable to the off-the-shelf He-Ne’s, the higher output intensity gives an opportunity i.e. to deliver enough power to drive multiple measurement axes with a single source. Moreover the DBR diode laser optical frequency stabilization to a selected molecular iodine transition gives advantage in traceability of the laser wavelength and thus the displacement measuring interferometeric system itself.

%
%
\section{Conclusions}

We have assembled an experimental laser system, based on a narrow-linewidth DBR laser diode lasing at the wavelength of $633$ nanometres. 
We have experimentally demonstrated a relative stability of $0.85 \times 10^{-9}$ over~$1$ minute and $1.65 \times 10^{-9}$ over~$1$~hour and absolute frequency reproducibility $\sigma = 7.1\,$MHz. 
The laser has a slightly worse degree of polarization ($1:10$) and broader linewidth ($1.2\,$MHz) compared to a typical He-Ne laser. 
The measurement on the optical bench (with two interferometers fed from the two laser sources) revealed a good coincidence between the tested and reference interferometer: the measurements at a individual points exhibited the stability in order of $2.5 \times 10^{-9}$. The systematic errors on the absolute scale were below $30\,$nm through the measurement range of $20\,$mm. The laser system exhibited the $1\,$nm tuning range with the temperature tuning and modulation bandwidth of several MHz with the injection current modulation.

These results indicate that the concept of a laser source represents a feasible alternative to frequency stabilized He-Ne lasers in dimensional (nano)metrology, especially where the traceability, high output power and also the the wide-band mode-hop free wavelength tuning options can bring significant benefits, i.e. in multi-axis displacement metrology, absolute homodyne interferometry. Together with the optical frequency modulation based detection methods \cite{Sensors12b} it also represents an entirely new concept in the dimensional metrology based on laser interferometry.  

We expect that our future effort will lead towards further challenges such as improvement of coupling efficiency, incorporation of the intensity stabilization, investigation of a wider range of frequency stabilization methods or achieving a more compact form factor. 

%
%

\section*{Acknowledgement}

The authors acknowledge the support from EMRP REG1 project IND58 6DoF (The EMRP is jointly funded by the EMRP participating countries within EURAMET and the European Union), from Academy of Sciences of the Czech Republic project RVO: 68081731.
The issue of optical frequency stabilization has been supported by Grant Agency of the Czech Republic (project GB14-36681G). 
The infrastructure has been supported from Ministry of Education, Youth and Sports of the Czech Republic (LO1212) together with the European Commission (ALISI No. CZ.1.05/2.1.00/01.0017). The construction of the laser head module has been funded by Technology Agency of the Czech Republic (project TE01020233).
NPL work was funded through National Measurement Programme for Engineering measurement.

The authors would also like to express their appreciation to John Mountford (NPL) as well as to Martin Sarbort, Jan Hrabina, Bretislav Mikel, Mirka Hola, Lenka Pravdova and Adam Lesundak (ISI) for secondment, Jens Flügge (PTB) for providing the laser diode and NPL workshop for custom mechanical parts. Simon Rerucha thanks NPL for hosting his secondment.

\subsection*{Disclaimer} The naming of any manufacturer or supplier by NPL or ISI in this article shall not be taken to be either NPL's or ISI's endorsement of specific samples of products of the said manufacturer; or recommendation of the said supplier. Furthermore, NPL and ISI cannot be held responsible for the use of, or inability to use, any products mentioned herein that have been used by NPL or ISI.

%
%

\section*{References}
\label{references}

\bibliographystyle{unsrt}   
\bibliography{dbr633}   

\end{document}